\documentclass{jetpl}
\twocolumn

\lat

\title{``Destruction'' of the Fermi surface due to pseudogap fluctuations
in strongly correlated systems}

\rtitle{``Destruction'' of the Fermi surface}

\sodtitle{``Destruction'' of the Fermi surface due to pseudogap fluctuations
in strongly correlated systems}

\author{E.\,Z.\,Kuchinskii$^{+}$, I.\,A.\,Nekrasov$^{+}$,
M.\,V.\,Sadovskii$^+$\/\thanks{e-mail: sadovski@iep.uran.ru}}

\rauthor{E.\,Z.\,Kuchinskii E.Z., I.\ A.\ Nekrasov I.A.,M.\,V.\,Sadovskii M.V.}

\sodauthor{Kuchinskii, Nekrasov, Sadovskii }

\address{$^+$Institute for Electrophysics, RAS Ural Branch,
620016 Ekaterinburg, Russia}

\dates{9 June 2005}{*}

\abstract{
We generalize the dynamical--mean field theory (DMFT) 
by including into the DMFT equations dependence on correlation length of 
pseudogap fluctuations via additional (momentum dependent) self--energy 
$\Sigma_{\bf k}$.  This self -- energy describes non-local dynamical 
correlations induced by short--ranged collective SDW--like antiferromagnetic 
spin (or CDW--like charge) fluctuations.  At high enough temperatures these 
fluctuations can be viewed as a quenched Gaussian random field with finite 
correlation length.  This generalized DMFT+$\Sigma_{\bf k}$ approach is used 
for the numerical solution of the weakly doped one--band Hubbard model with 
repulsive Coulomb interaction on a square lattice with nearest and next 
nearest neighbour hopping.  The effective single impurity problem 
is solved by numerical renormalization group (NRG).  
Both types of strongly correlated metals, namely (i) doped Mott 
insulator and (ii) the case of bandwidth $W\lesssim U$ ($U$ --- value of 
local Coulomb interaction) are considered. 
Calculating profiles of spectral densities for different parameters of the
model we demonstrate the qualitative picture of Fermi surface ``destruction'' 
and formation of ``Fermi arcs'' due to pseudogap fluctuations in qualitative
agreement with ARPES experiments. ``Blurring'' of the Fermi surface is 
enhanced with the growth of the Coulomb interaction.

}


\begin{document}

\maketitle

Pseudogap formation in the electronic spectrum of underdoped copper oxides 
\cite{Tim,MS} is is especially striking anomaly of the normal state of
high temperature superconductors.  
Despite continuing discussions on the nature of
the pseudogap, we believe that the preferable ``scenario'' for its formation 
is most likely based on the model of strong scattering of the charge
carriers by short--ranged antiferromagnetic (AFM, SDW) spin fluctuations
\cite{MS,Pines}. In momentum representation this scattering transfers 
momenta of the order of ${\bf Q}=(\frac{\pi}{a},\frac{\pi}{a})$ 
($a$ --- lattice constant of two dimensional lattice). 
This leads to the formation of structures in the one-particle spectrum, 
which are precursors of the changes in the spectra due
to long--range AFM order (period doubling).
As a result we obtain non--Fermi liquid like behavior (dielectrization)
of the spectral density in the vicinity of the so called ``hot spots'' on the
Fermi surface, appearing at intersections of the Fermi surface 
with antiferromagnetic Brillouin zone boundary \cite{MS}.

Within this spin--fluctuation scenario a simplified model of the pseudogap 
state was studied \cite{MS,Sch,KS} under the assumption that the scattering
by dynamic spin fluctuations can be reduced for high enough temperatures
to a static Gaussian random field (quenched disorder) of pseudogap fluctuations.
These fluctuations are characterized by a scattering vector from the 
vicinity of ${\bf Q}$,  with a width determined by the inverse correlation 
length of short--range order $\kappa=\xi^{-1}$, and by appropriate energy
scale $\Delta$ (typically of the order of crossover temperature $T^*$ to 
the pseudogap state \cite{MS}). 

Undoped cuprates are antiferromagnetic Mott insulators with
$U\gg W$ ($U$ --- value of local Coulomb interaction, $W$ --- bandwidth of
non--interacting band), so that correlation effects are actually very important.  
It is thus clear that the electronic properties of 
underdoped (and probably also optimally doped) cuprates are governed by 
strong electronic correlations too, so that these systems are typical 
strongly correlated metals. Two types of correlated metals can be 
distinguished:  (i) the doped Mott insulator and (ii) the bandwidth 
controlled correlated metal $W\approx U$. 

A state of the art tool to describe such correlated  systems
is the dynamical mean--field theory (DMFT)
\cite{MetzVoll89,vollha93,pruschke,georges96,PT}.
The characteristic features of correlated systems within the DMFT
are the formation of incoherent structures, the so-called Hubbard bands,
split by the Coulomb interaction $U$, and a
quasiparticle (conduction) band near the Fermi level dynamically 
generated by the local correlations 
\cite{MetzVoll89,vollha93,pruschke,georges96,PT}.

Unfortunately, the DMFT is not useful to the study the
``antiferromagnetic'' scenario of pseudogap formation in strongly
correlated metals. This is due to the basic approximation of the DMFT, which
amounts to the complete neglect of non-local dynamical correlation effects
\cite{MetzVoll89,vollha93,pruschke,georges96,PT}. As a result, within the
standard DMFT approach Fermi surface of a quasiparticle band is not 
renormalized by interactions and just coincides with that of the ``bare'' 
quasiparticles \cite{vollha93}.  Recently we have formulated a 
semiphenomenological DMFT+$\Sigma_{\bf k}$ approach \cite{cm05}, allowing the
introduction of a length scale (non--local correlations) into DMFT. 
Below we present basic points of this approach with application to the Fermi 
surface renormalization due to pseudogap fluctuations.

To include non--local effects, while remaining within the usual ``impurity
analogy''of DMFT, we propose the following procedure. To be definite, 
let us consider a standard one-band Hubbard model. 
The major assumption of our approach is that the lattice
and Matsubara ``time'' Fourier transform of the single-particle Green function 
can be written as:
\begin{equation}
G_{\bf k}(\omega)=\frac{1}{i\omega+\mu-\varepsilon({\bf k})-\Sigma(\omega)
-\Sigma_{\bf k}(\omega)}
\label{Gk}
\end{equation}
where $\Sigma(\omega)$ is the {\em local} contribution to self--energy,
surviving in the DMFT $(\omega=\pi T(2n+1))$, while $\Sigma_{\bf k}(\omega)$
is some momentum dependent part. We suppose that
this last contribution is due to either electron interactions with some
``additional'' collective modes or order parameter fluctuations, or may be
due to similar non--local contributions within the Hubbard model itself. 
To avoid possible confusion we must stress that $\Sigma_{\bf k}(i\omega)$
can also contain  local (momentum independent) contribution which obviously
{\em vanishes} in the limit of infinite dimensionality $d\to\infty$ and is
not taken into account within the standard DMFT. 
Due to this fact there is no double counting problem within our approach for 
the Hubbard model. It is important to stress that the 
assumed additive form of self--energy $\Sigma(\omega)+\Sigma_{\bf k}(\omega)$
implicitly corresponds to neglect of possible interference
of these local (DMFT) and non--local contributions.

The self--consistency equations of our generalized DMFT+$\Sigma_{\bf k}$ 
approach are formulated as follows \cite{cm05}:
\begin{enumerate}
\item{Start with some initial guess of {\em local} self--energy
$\Sigma(\omega)$, e.g. $\Sigma(\omega)=0$}.  
\item{Construct $\Sigma_{\bf k}(\omega)$ within some (approximate) scheme, 
taking into account interactions with collective modes or order parameter
fluctuations which in general can depend on $\Sigma(\omega)$
and $\mu$.} 
\item{Calculate the local Green function  
\begin{equation}
G_{ii}(\omega)=\frac{1}{N}\sum_{\bf k}\frac{1}{i\omega+\mu
-\varepsilon({\bf k})-\Sigma(\omega)-\Sigma_{\bf k}(\omega)}.
\label{Gloc}
\end{equation}
}
\item{Define the ``Weiss field''
\begin{equation}
{\cal G}^{-1}_0(\omega)=\Sigma(\omega)+G^{-1}_{ii}(\omega).
\label{Wss}
\end{equation}
}
\item{Using some ``impurity solver'' to calculate the single-particle Green function 
for the  effective Anderson impurity problem, defined by Grassmanian integral 
\begin{equation}
G_{d}(\tau-\tau')=\frac{1}{Z_{\text{eff}}}
\int Dc^+_{i\sigma}Dc_{i\sigma}
c_{i\sigma}(\tau)c^+_{i\sigma}(\tau')\exp(-S_{\text{eff}})
\label{AndImp}
\end{equation}
with effective action for a fixed (``impurity'') $i$
\begin{eqnarray}
S_{\text{eff}}=-\int_{0}^{\beta}d\tau_1\int_{0}^{\beta}
d\tau_2c_{i\sigma}(\tau_1){\cal G}^{-1}_0(\tau_1-\tau_2)c^+_{i\sigma}(\tau_2)+
\nonumber\\
+\int_{0}^{\beta}d\tau Un_{i\uparrow}(\tau)n_{i\downarrow}(\tau)\nonumber\\
\label{Seff}
\end{eqnarray}
$Z_{\text{eff}}=\int Dc^+_{i\sigma}Dc_{i\sigma}\exp(-S_{\text{eff}})$, and
$\beta=T^{-1}$. 
This step produces a {\em new} set of values $G^{-1}_{d}(\omega)$.}
\item{Define a {\em new} local self--energy
\begin{equation}
\Sigma(\omega)={\cal G}^{-1}_0(\omega)-
G^{-1}_{d}(\omega).
\label{StS}
\end{equation}
}
\item{Using this self--energy as ``initial'' one in step 1, continue 
the procedure until (and if) convergency is reached to obtain
\begin{equation}
G_{ii}(\omega)=G_{d}(\omega).
\label{G00}
\end{equation}
}
\end{enumerate}
Eventually, we get the desired Green function in the form of (\ref{Gk}),
where $\Sigma(\omega)$ and $\Sigma_{\bf k}(\omega)$ are those appearing
at the end of our iteration procedure.

For the momentum dependent part of the single-particle self--energy we 
concentrate on the effects of scattering of electrons from collective 
short-range SDW--like antiferromagnetic spin (or CDW--like charge) fluctuations.
To calculate $\Sigma_{\bf k}(\omega)$ for an electron moving in the quenched
random field of (static) Gaussian spin (or charge) fluctuations with dominant
scattering momentum transfers from the vicinity of some characteristic
vector ${\bf Q}$  (``hot spots'' model \cite{MS}), 
we use the following recursion procedure 
proposed in Refs.~\cite{MS79,Sch,KS} which takes into account {\em all} 
Feynman diagrams describing the scattering of electrons by this random field:  
\begin{equation}
\Sigma_{\bf k}(\omega)=\Sigma_{n=1}(\omega{\bf k})
\label{Sk}
\end{equation}
with
\begin{equation}
\Sigma_{n}(\omega{\bf k})=
\Delta^2\frac{s(n)}{i\omega+\mu-\Sigma(\omega)
-\varepsilon_n({\bf k})+inv_n\kappa-\Sigma_{n+1}(\omega{\bf k})}\\ 
\label{rec}
\end{equation} 
The quantity $\Delta$ characterizes the energy scale and
$\kappa=\xi^{-1}$ is the inverse correlation length of short range
SDW (CDW) fluctuations, $\varepsilon_n({\bf k})=\varepsilon({\bf k+Q})$ and 
$v_n=|v_{\bf k+Q}^{x}|+|v_{\bf k+Q}^{y}|$ 
for odd $n$ while $\varepsilon_n({\bf k})=\varepsilon({\bf k})$ and $v_{n}=
|v_{\bf k}^x|+|v_{\bf k}^{y}|$ for even $n$. The velocity projections
$v_{\bf k}^{x}$ and $v_{\bf k}^{y}$ are determined by usual momentum derivatives
of the ``bare'' electronic energy dispersion $\varepsilon({\bf k})$. Finally,
$s(n)$ represents a combinatorial factor with
\begin{equation}
s(n)=n
\label{vcomm}
\end{equation}
for the case of commensurate charge (CDW type) fluctuations with
${\bf Q}=(\pi/a,\pi/a)$ \cite{MS79}. 
For incommensurate CDW fluctuations \cite{MS79} one finds
\begin{equation} 
s(n)=\left\{\begin{array}{cc}
\frac{n+1}{2} & \mbox{for odd $n$} \\
\frac{n}{2} & \mbox{for even $n$}.
\end{array} \right.
\label{vinc}
\end{equation}
If we take into account the (Heisenberg) spin structure of interaction with 
spin fluctuations in  ``nearly antiferromagnetic Fermi--liquid'' 
(spin--fermion (SF) model Ref.~\cite{Sch}),
combinatorics of diagrams becomes more complicated and 
factor $s(n)$ acquires the following form \cite{Sch}:
\begin{equation} 
s(n)=\left\{\begin{array}{cc}
\frac{n+2}{3} & \mbox{for odd $n$} \\
\frac{n}{3} & \mbox{for even $n$}.
\end{array} \right.
\label{vspin}
\end{equation}
Obviously, with this procedure we introduce an important length scale $\xi$ 
not present in standard DMFT. Physically this scale mimics the effect of 
short--range (SDW or CDW) correlations within fermionic ``bath'' surrounding 
the effective Anderson impurity. Both parameters $\Delta$ and $\xi$ can in
principle be calculated from the microscopic model at hand \cite{cm05}. 

In the following we will consider 
both $\Delta$ and especially $\xi$ as some phenomenological parameters to be 
determined from experiments. This makes our approach somehow similar
in the spirit to Landau approach to Fermi--liquids.

In the following, we discuss a standard one-band
Hubbard model on a square lattice. With nearest ($t$) and
next nearest ($t'$) neighbour hopping integrals the ``bare'' dispersion
then reads
\begin{equation}
\varepsilon({\bf k})=-2t(\cos k_xa+\cos k_ya)-4t'\cos k_xa\cos k_ya\;\;,
\label{spectr}
\end{equation}
where $a$ is the lattice constant.
The correlations are introduced by a repulsive local two-particle interaction 
$U$. We choose as energy scale the nearest neighbour hopping integral $t$
and as length scale the lattice constant $a$. All energies below are given in 
units of $t$.

For a square lattice the bare bandwidth is $W=8t$.
To study a strongly correlated metallic state obtained as doped Mott insulator
we use $U=40t$ as value for the Coulomb interaction and a filling $n=0.8$ (hole 
doping).  The correlated metal in the case of $W\gtrsim U$ 
is considered for the case of $U=4t$ and filling factor $n=0.8$ 
(hole doping). For $\Delta$ we choose rather typical values between 
$\Delta=0.1t$ and $\Delta=2t$ (actually as approximate limiting values 
obtained in Ref.\cite{cm05}) and for the 
correlation length we have taken $\xi=10a$ (being 
motivated mainly by experimental data for cuprates~\cite{MS,Sch}).

The DMFT maps the lattice problem onto an effective, 
self--consistent impurity defined by Eqs. (\ref{AndImp})-(\ref{Seff}).  
In our work we employed as ``impurity solver'' a reliable method of
numerical renormalization group (NRG) \cite{NRG,BPH}.

As already discussed in the Introduction, the characteristic feature of the 
strongly correlated metallic state is the coexistence of lower and upper 
Hubbard bands split by the value of $U$ with a quasiparticle peak at the Fermi 
level.

Once we get a self--consistent solution of the DMFT+$\Sigma_{\bf k}$ 
equations with non-local fluctuations we can compute the spectral functions
$A(\omega,{\bf k})$ for real $\omega$:
\begin{equation}
A(\omega,{\bf k})=-\frac{1}{\pi}{\rm Im}\frac{1}{\omega+\mu
-\varepsilon({\bf k})-\Sigma(\omega)-\Sigma_{\bf k}(\omega)},
\label{specf}
\end{equation}
where self--energy $\Sigma(\omega)$ and chemical potential $\mu$
are calculated self--consistently. Densities of states can be calculated 
integrating (\ref{specf}) over the Brillouin zone.

Extensive calculations of the densities of states, spectral densities and
ARPES spectra for this model were performed in Ref. \cite{cm05}.
In general case pseudogap appears in the density of states appears within the
in the quasiparticle peak (correlated conduction band). 
Qualitative behaviour of the pseudogap anomalies is similar to those
for the case of $U=0$ \cite{MS,KS}, e.g.\ a decrease of $\xi$ makes the 
pseudogap less pronounced, while reducing $\Delta$ narrows the pseudogap 
and also makes it more shallow. 
For the doped Mott--insulator we find that the pseudogap is 
remarkably more pronounced for the SDW--like fluctuations than for CDW--like 
fluctuations. Thus below we present mainly the results obtained using
combinatorics (\ref{vspin}) of spin -- fermion model.

As was noted above within the standard DMFT approach Fermi surface is not 
renormalized by interactions and just coincides with that of the ``bare'' 
quasiparticles \cite{vollha93}. However, in the case of nontrivial momentum 
dependence of electron self -- energy, important renormalization of the Fermi 
surface appears due to pseudogap formation \cite{Sch}.  
There are a number of ways to define Fermi surface in strongly correlated 
system with pseudogap fluctuations. 
In the following we are using intensity plots (within the 
Brillouin zone) of the spectral density (\ref{specf}) taken at 
$\omega=0$. These are readily measured by ARPES and appropriate peak 
positions define the Fermi surface in the usual Fermi liquid case. 

\begin{figure}
\includegraphics[clip=true,width=0.5\textwidth]{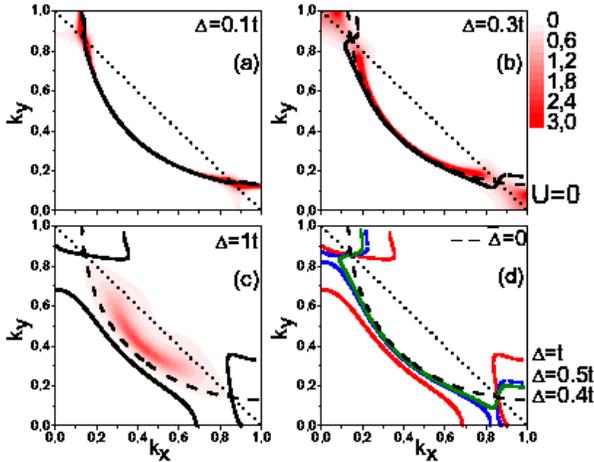}
\caption{Fig.1. Fermi surfaces obtained for uncorrelated case 
of $U=0$ and filling factor $n=0.8$.
Shown are intensity plots of spectral density (\ref{specf}) for 
$\omega=0$. 
(a) -- $\Delta=0.1t$;\ (b) -- $\Delta=0.3t$;\ (c) -- $\Delta=t$;\
(d) -- ``Fermi surfaces'' obtained solving Eq. (\ref{ReFS}).
Dashed line denotes ``bare'' Fermi surface.}
\label{FS_U0}
\end{figure}

In Fig. \ref{FS_U0} (a-c) we show such plots for the case of uncorrelated metal
($U=0$) with pseudogap fluctuations, obtained directly from the Green's
function defined by the recursion procedure (\ref{Sk}), (\ref{rec}).
For comparison, in Fig. \ref{FS_U0} (d) we show renormalized Fermi surfaces
obtained for this model using rather natural definition
of the Fermi surface as defined by the solution of the equation
\begin{equation}
\omega-\varepsilon({\bf k})+\mu-Re\Sigma(\omega)-Re\Sigma_{\bf k}(\omega)=0
\label{ReFS}
\end{equation}
for $\omega=0$, used e.g. in Ref.\cite{Sch}. It is seen, that this last
definition produces the Fermi surfaces close to those defined by intensity
plots of the spectral density only for small values of $\Delta$, while for
larger values we can see rather unexpected ``topological transition''.
At the same time, spectral density intensity plots clearly demonstrate 
``destruction'' of the Fermi surface in the vicinity of the ``hot spots'' 
with ``Fermi arcs'' formation with the growth of $\Delta$, similar to those 
seen in pioneering ARPES experiments of Norman et al. \cite{Norm}, and 
confirmed later in numerous works. 

\begin{figure}
\includegraphics[clip=true,width=0.5\textwidth]{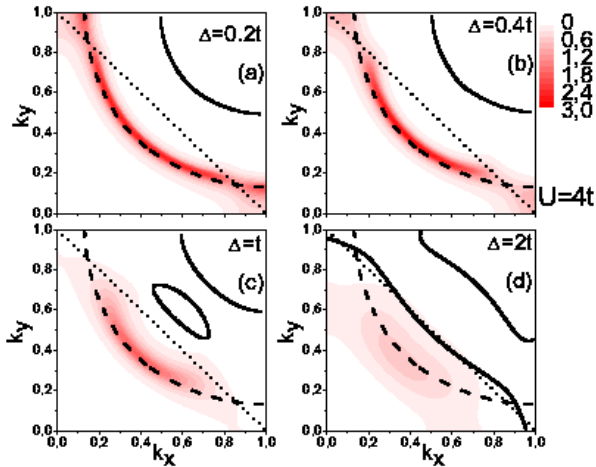}
\caption{Fig.2.``Destruction''of the Fermi surface as obtained from the
DMFT+$\Sigma_{\bf k}$ calculations for $U=4t$ and $n=0.8$.
Notations are the same as used in Fig. \ref{FS_U0}.
(a) -- $\Delta=0.2t$;\ (b) -- $\Delta=0.4t$;\ (c) -- $\Delta=t$;\
(d) -- $\Delta=2t$.
Black lines show the solution of Eq. (\ref{ReFS}).}
\label{FS_U4}
\end{figure}

\begin{figure}
\includegraphics[clip=true,width=0.5\textwidth]{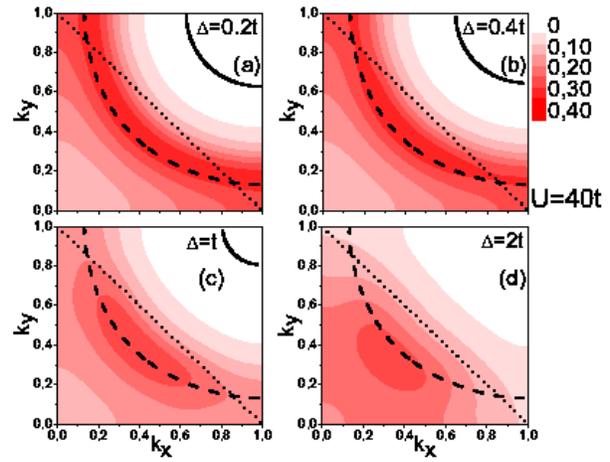}
\caption{Fig.3. ``Fermi surfaces'' obtained from the
DMFT+$\Sigma_{\bf k}$ calculations for $U=40t$ and $n=0.8$.
Other parameters and notations are the same as in Fig.~\ref{FS_U4}.}
\label{FS_U40}
\end{figure}

In Fig.~\ref{FS_U4} we show our results for the case of correlated metal
with $U=4t$ and in Fig.~\ref{FS_U40} for the doped Mott insulator with $U=40t$. 
Again we see the qualitative behavior clearly demonstrating 
the ``destruction'' of the well defined Fermi surface in the strongly 
correlated metal with the growth of the pseudogap amplitude $\Delta$.  
The role of finite $U$ reduces to lower intensity of spectral density in
comparison with the case of $U=0$ and leads to additional ``blurring'', 
making ``hot spots'' less visible.  Again the ``destruction'' of the Fermi 
surface starts in the vicinity of ``hot spots'' for small values of $\Delta$, 
but almost immediately it disappears in the 
whole antinodal region of the Brillouin zone, while only ``Fermi arcs'' 
remain in the nodal region very close to the ``bare'' Fermi surface. These 
results give a natural explanation of the observed behavior and also of the 
fact that the existence of ``hot spots'' regions was observed only in some 
rare cases \cite{Arm}.  

For the case of doped Mott insulator ($U=40t$) shown in 
Fig.~\ref{FS_U40} we see that the ``Fermi surface'' is rather poorly defined 
for all values of $\Delta$, as the spectral density profiles are much more
``blurred'' than in the case of smaller values of $U$, reflecting important
role of correlations. 

It is interesting to note that from 
Figs.~\ref{FS_U4}, \ref{FS_U40} it is clearly seen that the ``natural'' 
definition of the Fermi surface from Eq. (\ref{ReFS}) is quite inadequate 
for correlated systems with finite $U$ and nonlocal interactions (pseudogap 
fluctuations), signifying the increased role of strong correlations. 

To summarize, we propose a generalized DMFT+$\Sigma_{\bf k}$
approach, which is meant to take into account the important
effects due to non--local correlations in a systematic, but to some extent
phenomenological fashion.
The main idea of this extension is to stay within a
usual effective Anderson impurity analogy, and introduce length scale
dependence due to non-local correlation via the effective medium (``bath'') 
appearing in the standard DMFT. 
This becomes possible by incorporating scattering processes of fermions
in the ``bath'' from non-local collective SDW--like antiferromagnetic spin (or
CDW--like charge) fluctuations.
Such a generalization of the DMFT allows one to overcome the well--known 
shortcoming of ${\bf k}$--independence of self--energy of the standard DMFT.  
It in turn opens the possibility to access the physics of low--dimensional 
strongly correlated systems, where different types of spatial fluctuations 
(e.g. of some order parameter), become important in a non-perturbative way at
least with respect to the important local dynamical correlations. 
However, we must stress that our procedure in no way introduces any kind of
systematic $1/d$--expansion, being only a qualitative method to include
length scale into DMFT.

In our present study we addressed the problem of the Fermi surface
renormalization (``destruction'') by pseudogap fluctuations in the
strongly correlated metallic state.  
Our generalization of DMFT leads to non--trivial 
and in our opinion physically sensible ${\bf k}$--dependence of spectral 
functions, leading to Fermi surface renormalization quite similar to that
observed in ARPES experiments. 

Similar results were obtained in recent years using the cluster mean-field 
theories \cite{TMrmp}. The major advantage of our approach over these
cluster mean-field theories is, that we stay in an effective single-impurity 
picture. This means that our approach is computationally much less expensive 
and therefore easily generalizable for the account of additional interactions.

We are grateful to Th. Pruschke for providing us with his NRG code and
helpful discussions.
This work was supported in part by RFBR grants 05-02-16301, 05-02-17244, 
and programs of the
Presidium of the Russian Academy of Sciences (RAS) ``Quantum macrophysics''
and of the Division of Physical Sciences of the RAS ``Strongly correlated
electrons in semiconductors, metals, superconductors and magnetic
materials''. I.N. acknowledges support
from the Dynasty Foundation and International
Centre for Fundamental Physics in Moscow program for young
scientists 2005 and Russian Science Support Foundation program for
young PhD of the Russian Academy of Sciences 2005.

\pagestyle{empty}

\end{document}